\documentclass[conference]{IEEEtran}

\usepackage{amsmath,amssymb,amsfonts}
\usepackage{algorithmic}

\usepackage{graphicx}
\usepackage{textcomp}
\usepackage{xcolor}
\usepackage{tikz}
\usepackage{enumitem}

\usepackage{soul}
\definecolor{light-yellow}{RGB}{254, 247, 178}
\definecolor{light-orange}{RGB}{245, 225, 187}
\definecolor{light-red}{RGB}{227, 189, 189}
\definecolor{light-purple}{rgb}{0.902, 0.694, 0.831}

\usepackage{booktabs}
\usepackage{tabularx}

\usepackage{cite}
\usepackage{hyperref}
\usepackage{cleveref}

\def\BibTeX{{\rm B\kern-.05em{\sc i\kern-.025em b}\kern-.08em
    T\kern-.1667em\lower.7ex\hbox{E}\kern-.125emX}}

\newcommand \copyrighttext {
  \footnotesize \textcopyright 2025 IEEE. Personal use of this material is permitted. Permission from IEEE must be obtained for all other uses, in any current or future media, including reprinting/republishing this material for advertising or promotional purposes, creating new collective works, for resale or redistribution to servers or lists, or reuse of any copyrighted component of this work in other works. DOI: (to be added once issued) 
}

\newcommand\copyrightnotice{
    \begin{tikzpicture}[remember picture,overlay]
    \node[anchor=south,yshift=10pt] at (current page.south) {\fbox{\parbox{\dimexpr\textwidth-\fboxsep-\fboxrule\relax}{\copyrighttext}}};
    \end{tikzpicture}
}
    
\begin{document}

\title{Adopting Use Case Descriptions for Requirements Specification: an Industrial Case Study}

\author{
    \IEEEauthorblockN{
        Julian Frattini}
    \IEEEauthorblockA{\textit{Chalmers University of Technology}\\
        \textit{and University of Gothenburg}\\
        Gothenburg, Sweden \\
        julian.frattini@chalmers.se}
    \and
    \IEEEauthorblockN{
        Anja Frattini}
    \IEEEauthorblockA{\textit{FernUniversität in Hagen}\\
        Hagen, Germany}
}

\maketitle
\copyrightnotice

\begin{abstract}
    \textbf{Context}: 
    Use case (UC) descriptions are a prominent format for specifying functional requirements.
    Existing literature abounds with recommendations on how to write high-quality UC descriptions but lacks insights into (1) their real-world adoption, (2) whether these recommendations correspond to actual quality, and (3) which factors influence the quality of UCs.
    \textbf{Objectives}:
    We aim to contribute empirical evidence about the adoption of UC descriptions in a large, globally distributed case company.
    \textbf{Methods}: 
    We surveyed 1188 business requirements of a case company that were elicited from 2020-01-01 until 2024-12-31 and contained 1192 UCs in various forms.
    Among these, we manually evaluated the 273 template-style UC descriptions against established quality guidelines.
    We generated descriptive statistics of the format's adoption over the surveyed time frame.
    Furthermore, we used inferential statistics to determine (a) how properties of the requirements engineering process affected the UC quality and (b) how UC quality affects subsequent software development activities.
    \textbf{Results and Conclusions}: 
    Our descriptive results show how the adoption of UC descriptions in practice deviates from textbook recommendations.
    However, our inferential results suggest that only a few phenomena like solution-orientation show an actual impact in practice.
    These results can steer UC quality research into a more relevant direction.
\end{abstract}


\section{Introduction}
\label{sec:intro}

Requirements artifacts are essential work products from the requirements engineering (RE) process~\cite{mendez2013improving,mendez2019artefacts}.
They specify needs and constraints as well as desired system properties which relevant stakeholders impose on the system under development~\cite{glinz2011glossary}.
Their importance for the remaining software engineering (SE) life cycle stems from the fact that requirements artifacts inform subsequent activities performed by different roles~\cite{frattini2024measuring}.
For example, software architects infer a system architecture from non-functional requirements, and developers implement features described in functional requirements.

Given the relevance of requirements artifacts, an adequate format to specify a requirement is essential~\cite{mendez2015artefact,mendez2010meta}.
One common format is the \textit{use case (UC) description} introduced by Ivar Jacobson~\cite{jacobson1993object}.
Its clear structure~\cite{cockburn1998basic} and purpose~\cite{jacobson1993object} popularized it for specifying functional requirements.

As with all requirements artifacts, the quality of UC descriptions impacts how well subsequent activities which use them perform~\cite{femmer2018requirements,frattini2023requirements}.
Consequently, prior research proposed quality criteria for writing ``good'' UCs~\cite{phalp2007assessing,adolph2002patterns,usdadiya2019empirical}.
However, literature lacks empirical insights into the adoption and application of UCs in practice, but also whether the proposed quality criteria actually support the usability of the artifact. 
We argue that both are fundamental for problem-oriented research and to eventually support the SE industry in deciding whether and how to adopt the UC description format.

\begin{figure}
    \centering
    \includegraphics[width=1\linewidth]{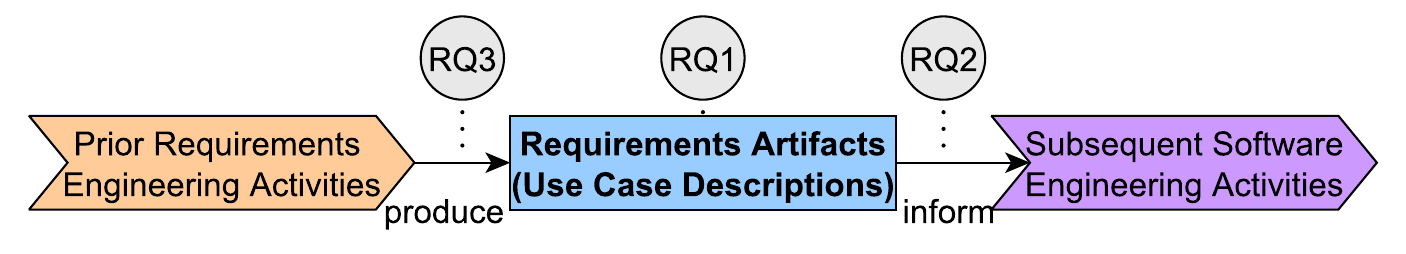}
    \caption{Mapping of RQs to the studied Context}
    \label{fig:rqs}
\end{figure}

To address this gap, we conducted a case study at a large software company in Sweden were we investigated the adoption of UC descriptions, the impact of their quality, and factors influencing this quality.
We investigated the following research questions which are also visualized in the outlined context in \Cref{fig:rqs}:

\begin{itemize}
    \item RQ1: How are UC descriptions adopted at the case company?
    \item RQ2: How does UC quality impact the subsequent solution design phase?
    \item RQ3: Which RE factors influence the UC quality?
\end{itemize}

Our case study contributes insights into the real-world application of UC descriptions for requirements specification.
The inferential analysis reveals phenomena of causal effects among the RE process, UC quality, and SE activities.
These indicate, which properties of UC quality have an actual impact in practice and deserve further attention.

The rest of the paper is structured as follows.
\Cref{sec:related} introduces related work on UCs and requirements quality.
\Cref{sec:method} presents our research method and \Cref{sec:results} the obtained results.
We discuss these in \Cref{sec:discussion} before concluding in \Cref{sec:conclusion}.

\subsection*{Data Availability Statement}

We share all material generated in the scope of this study to the extent possible at \url{https://doi.org/10.5281/zenodo.15672950}.
We cannot share any of the sensitive raw data beyond the examples presented later in the manuscript.
Therefore, we supplement our replication package with mocked data that allows to execute all study scripts to ease its reuse.

\section{Related Work}
\label{sec:related}


\subsection{Use Cases}
\label{sec:related:uc}

A UC is a ``collection of possible sequences of interactions between the system under discussion and its external actors, related to a particular goal''~\cite{cockburn1997structuring}.
In his seminal book~\cite{jacobson1993object}, Jacobson introduced the UC-driven approach that produces a \textit{UC model}, i.e., a combination of a high-level UC overview diagram and UC descriptions.


The format in which a UC descirption may be specified can vary.
While Nardi originally recommended writing ``one or two sides of well-crafted prose''~\cite{nardi1992use}, Polley and Stevens~\cite{stevens2006using} just as Rosenberg~\cite{rosenberg1999use} simply propose writing UCs as paragraphs.
Finally, Harwood suggested a tabular format~\cite{harwood1997use} which Cockburn popularizes in his seminal book on ``Writing Effective Use Cases''~\cite{cockburn2008writing}.
This tabular ``template''-style specifies attributes like the involved actors, pre- and post-conditions, and, most importantly, a main scenario extended by one or more alternative scenarios written as sequences of interactions between the actors and the system.
\Cref{fig:uc} shows a simple example of a template-style UC description with highlighted defects that we explain later.
Their clear structure promoted them to a popular format for specifying functional requirements~\cite{anda2003empirical}.


\begin{figure}
    \centering
    \begin{tikzpicture}
        \node[draw, text width=0.97\columnwidth, align=left](req){
            \small
            \textbf{Title}: Change subscription plan \\
            \vspace{0.1cm}
            \textbf{Actors}: End-user \\
            \vspace{0.1cm}
            \textbf{Preconditions}: The end-user has a valid and active subscription \\
            \vspace{0.1cm}
            \textbf{Main scenario}:
            \begin{enumerate}[label=\arabic*.]
                \item A user selects the ``upgrade account''-option.
                \item Once a new \sethlcolor{light-purple}\hl{subscription plan} \sethlcolor{light-yellow}\hl{is selected and confirmed}, the \sethlcolor{light-purple}\hl{UI} requests to select a \sethlcolor{light-purple}\hl{payment option}.
                \item If the user already has a valid payment option, they can select it. \sethlcolor{light-orange}\hl{If not, they must enter a new payment option.}
                \item \sethlcolor{light-red}\hl{The system updates the subscription plan in the user database and logs the change with a timestamp.}
            \end{enumerate}
        };
    \end{tikzpicture}

    \caption{Exemplary template-style UC description containing the defects \sethlcolor{light-purple}\hl{incoherent sentence}, \sethlcolor{light-yellow}\hl{passive voice}, \sethlcolor{light-orange}\hl{misplaced alternative}, and \sethlcolor{light-red}\hl{white-box step}.}
    \label{fig:uc}
\end{figure}

From the popularity of UCs, derivatives on several levels of abstraction emerged~\cite{cockburn2008writing}.
User-level UCs describe ``outwardly visible requirements of a system''~\cite{anda2003empirical} while system-level UCs provide details on system-internal communications.
In the scope of this work, we focus on the level relevant to RE, i.e., user-level UCs~\cite{kulak2001use}.

\subsection{Requirements Artifact Quality}
\label{sec:related:quality}

UC descriptions, like all requirements artifacts, inform subsequent SE activities~\cite{femmer2018requirements} and, therefore, the artifacts' quality impacts the performance of these activities~\cite{femmer2015s}, a phenomenon called activity-based requirements quality~\cite{frattini2023requirements}.
For example, a vague requirement artifact may lead to a developer misinterpreting a stakeholder's need and, therefore, develop a feature incorrectly.
Requirements quality research supports SE practitioners in identifying and mitigating these quality defects early~\cite{montgomery2022empirical} to avoid the cost for their removal increasing~\cite{boehm1984software}.
However, the current state of research focuses more on proposing new alleged quality defects but lacks empirical evidence on their impact~\cite{montgomery2022empirical,frattini2023requirements}.
As a consequence, many of these alleged defects are of questionable relevance to SE practitioners~\cite{frattini2022live,femmer2017rapid} .

UCs are no exception to this.
Several prior works have introduced quality criteria that allegedly correspond with high-quality UC descriptions~\cite{phalp2007assessing,adolph2002patterns,usdadiya2019empirical}.
These include meta-models for UC descriptions~\cite{siqueira2011essential}, writing guidelines~\cite{achour1999guiding}, quality metrics~\cite{usdadiya2019empirical}, and anti-patterns~\cite{el2012constructing}.
Phalp and Cox organized prior recommendations into a ``Use Case Description Quality Checklist''~\cite{phalp2007assessing} grounded in discourse processing research~\cite{traxler1995improving}.
This checklist consists of the following set of heuristics\footnote{Also named the `7 Cs' since all heuristics begin with a C} for higher quality UC descriptions~\cite{phalp2007assessing}:

\begin{enumerate}
    \item \textbf{Coverage}: a UC should contain precisely as much as necessary but as little as possible information to answer the problem
    \item \textbf{Cogent}: the scenarios in a UC should follow a logical, end-to-end path
    \item \textbf{Coherent}: each sentence should repeat a noun in the last sentence
    \item \textbf{Consistent abstraction}: all described interactions should be on a consistent level of abstraction (e.g., user-system communications)
    \item \textbf{Consistent structure}: alternative paths should be excluded from the main scenario and the numbering should be consistent
    \item \textbf{Consistent grammar}: sentences should be written in simple present tense and active voice
    \item \textbf{Consideration of alternatives}: alternative paths should be separate, viable, and numbered consistently with the main scenario
\end{enumerate}

\Cref{fig:uc} highlights four violations against these heuristics.
The use of \sethlcolor{light-yellow}\hl{passive voice} omits the actor from the second step.
Additionally, none of the three nouns in this step align with the two nouns from the first step.
These \sethlcolor{light-purple}\hl{incoherent} steps obscure whether ``upgrade account'' refers to changing a ``subscription plan.''
The \sethlcolor{light-orange}\hl{misplaced alternative} in the main scenario should be located in a separate section called ``alternatives'' or ``extensions.''
Finally, the \sethlcolor{light-red}\hl{white-box step} describes a system-system interaction which would be invisible to the actor of the UC and, therefore, irrelevant in a user-level UC.
Furthermore, this step imposes on the solution-space by determining a potentially non-optimal solution prior to fully specifying the problem~\cite{fernandez2012field}.



As previously mentioned, empirical evidence about the impact of quality criteria is sparse for UC descriptions.
Phalp et al. validated their proposed guidelines by applying them in two UCs at a case company and validating the detected defects with stakeholders, which subjectively support their importance~\cite{phalp2007assessing}.
Achour et al. conducted a controlled experiment comparing different variants of their CREWS writing guidelines~\cite{achour1999guiding}, but their positive results were later refuted by a replication from Cox and Phalp~\cite{cox2000replicating}.
Cox contributed a laboratory experiment on their own but focused more on evaluating the effectiveness of UC guidelines to support inspection techniques~\cite{cox2004experiment}, where they found that checklist-style guidelines bias an inspector to focus on syntactic aspects.
In another controlled experiment, Usdadiya et al. compared a proposed set of UC quality metrics with the subjective quality assessment of graduate students~\cite{usdadiya2019empirical}.
Anda et al. contributed a case study observing the changes that were made in UC descriptions~\cite{anda2003empirical}, yet without any empirical link between these changes and their quality.
As most empirical evidence is constrained to lab-settings, systematic assessments of the field agree that major gaps in literature remain the industrial relevance and empirical assessment of these quality guidelines~\cite{tiwari2015systematic}.

\section{Method}
\label{sec:method}

In this study, we start addressing the aforementioned gap of missing empirical evidence of real-world applications.
We conducted a case study to answer the research questions formulated in \Cref{sec:intro}.
Our study classifies as a case study since it is an empirical inquiry of a contemporary phenomenon (i.e., the current adoption of UC descriptions for requirements specification) in its real-life context~\cite{wohlin2021case}.
While case studies gain their realism only at the expense of control over the variables of interest, we deem this method necessary since the boundary between the phenomenon and the context, i.e., between UC description quality and its real-life use, is unclear~\cite{wohlin2021case}.
We designed and report our case study according to Runeson and H{\"o}st~\cite{runeson2009guidelines}.

\subsection{Case Study Plan}
\label{sec:method:case}

Our \textit{objective} is to describe the adoption of UC descriptions in practice (RQ1), to explain how their quality impacts subsequent SE activities (RQ2), and how the RE process impacts their quality (RQ3) as shown in \Cref{fig:rqs}.
This makes our case study both \textit{descriptive} (i.e., portraying a phenomenon in scope of RQ1) and \textit{explanatory} (i.e., explaining causal relationships in the scope of RQ2 and RQ3)~\cite{robson2002real}.
The \textit{theory} providing a frame of reference is the notion of activity-based requirements quality~\cite{frattini2023requirements,femmer2015s,femmer2018requirements} introduced in \Cref{sec:related:quality}.
The \textit{research questions} are stated in \Cref{sec:intro}.
We employ document analysis as a \textit{method}, supplemented with a focus group for validation~\cite{kontio2008focus}, and our \textit{selection strategy} is by availability~\cite{benbasat1987case}.
While the latter is not as purposive as it should be, selection by availability is common in SE research~\cite{runeson2009guidelines}.

Finally, our \textit{case} itself is the adoption of UC descriptions in a case company. 
The studied case was a Business Support System (BSS) for the telecommunication domain that manages critical business operations and customer-facing functions, provided by a large, globally distributed software development company originating from Sweden.
The system employs a microservice architecture and consists of 24 individual modules, each developed by their own sub-organization within the scope of our case.
Several of these modules provide customer-facing functionality of a BSS, including customer relation management, the management of orders, as well as billing and revenue management.
The released product based on this BSS consists of different configurations of these modules and is used worldwide.
We sampled the case by convenience~\cite{baltes2022sampling}, as it is managed at the site accessible to us and expressed an intrinsic interest in adopting a UC-driven approach.

The organization follows an agile SE paradigm with three-week sprints to incorporate customer feedback during the development period.
To accommodate the global scale of the development, the organization still relies on an overarching plan-driven structure divided into the following stages:

\begin{itemize}
    \item \textbf{Ideation} (S0): first abstract draft of a business requirement including a title and basic information (e.g., the assumed complexity level).
    \item \textbf{Requirements Engineering} (S1): decomposition of the business requirement into smaller requirements and agreement on their specification
    \item \textbf{Design} (S2): deriving a software solution specification from the requirements specification and finalizing the effort estimation
    \item \textbf{Implementation} (S3): implementation of the software solution based on the solution specification
    \item \textbf{Testing} (S4): verification and validation of the software solution based on the requirements and solution specification
    \item \textbf{Deployment} (S5): release of the product update
\end{itemize}

The organization uses Jira for requirements specification and management.
In Jira, requirements engineers specify business requirements particular to one or more of the aforementioned modules.
These user-level business requirements represent both requests for new features and updates of existing ones.
In 2020, the organization committed to adopt the UC format following the guidance by Cockburn~\cite{cockburn1998basic} to specify the decomposed, smaller requirements.
Additionally, business requirements record meta-information like the assumed complexity level or estimated effort, but also to track the development progress in terms of status and time spent per stage S0-S5.
Over the studied period, 30 different requirements engineers have authored requirements following a systematic, company-internal RE approach composed of several phases.
All phases are executed internally but assigned to globally distributed teams.

\subsection{Data Collection}
\label{sec:method:data}

\subsubsection{Data Acquisition}

The organization provided us access to their Jira system with a complete view on their requirements specifications.
We exported 1188 business requirements that completed at least S4 created between 2020-01-01 and 2024-12-31.
In addition to the title and description (i.e., the body of the requirement), each requirement contained the attributes listed in \Cref{tab:attributes:requirement}.

\begin{table}[hbt]
    \centering
    \caption{Attributes per business requirement}
    \label{tab:attributes:requirement}
    \begin{tabularx}{\linewidth}{lXp{1.7cm}}
    \toprule
        \textbf{Attribute} & \textbf{Description} & \textbf{Type} \\
        \midrule
        Owner & Product owner responsible for the requirement & acronym \\
        Author & Writer of the requirement & acronym \\
        Complexity & Assumed complexity of the requirement & [low, medium, high] \\
        S-dates & Start and end date of S0-S5 & date \\
        \bottomrule
    \end{tabularx}
\end{table}

\subsubsection{Manual Extraction of basic Attributes}

The description of each business requirement contained a varying number of UCs, where many requirements contained no UC at all but rather described their content in a free form.
We reviewed all 1188 requirements and manually extracted relevant information according to several attributes described in the following paragraphs.
The two authors distributed the attributes among each other and extracted them independently.
After each iteration of one week, we convened and discussed all decisions.
To ensure consistency and reliability of our manual extractions, we developed and documented a guideline containing instructions and examples.
We tested the guideline on 350 requirements ($350/1188=29.5\%$) before committing to the actual extraction on the full data set once the guidelines were stable.

We reviewed all 1188 requirements and first manually extracted the information specified in \Cref{tab:attributes:uc:basic} about each UC explicitly declared as such (i.e., prefixed by ``use case'' or ``uc'').
The \textit{form} attribute captures the variations in appearance in which use cases after explicit declaration appear.
Some UCs consisted only of a \textit{title}, others are written \textit{informally} as paragraphs of prose.
Some UCs actually employed a syntax resembling a \textit{user story} following the Connextra template~\cite{cohn2004user}.
The form of main interest were UC in a tabular \textit{template}-style.
\Cref{fig:uc} reflects an exemplary template-style UC.

We annotated the remaining attributes (fields and location) only for the UCs that had a template-style form, i.e., the only form to which the aforementioned attributes apply.
The \textit{fields} attribute lists all attributes contained in the UC, e.g., the title, actors, main scenario, and extensions.
Finally, the \textit{location} attribute captures, where the use case is located within the requirement.
If the description field of the business requirement follows the recommended document structure supplementing the ``Business Requirement'' with a ``Business Background'' and ``Solution Proposal'', we annotated in which section the UC was located.
If the description of the requirement did not follow the structure, we labeled the location as \textit{default}.

\begin{table}[hbt]
    \centering
    \caption{Basic attributes per use case}
    \label{tab:attributes:uc:basic}
    \begin{tabularx}{\linewidth}{lXp{3cm}}
    \toprule
        \textbf{Attribute} & \textbf{Description} & \textbf{Type} \\
        \midrule
        Form & Syntax used to specify the use case & [title, user story, template, informal] \\
        Fields & list of attributes in a use case description & list \\
        Location & Position of the use case within the requirement & [business background, business requirement, solution proposal, default] \\
        \bottomrule
    \end{tabularx}
\end{table}

\subsubsection{Manual Extraction of Quality Attributes}

We mainly aimed to evaluate the quality of template-style UCs.
For this, we annotated each UC according to the seven heuristics proposed by Phalp et al.~\cite{phalp2007assessing} presented in \Cref{sec:related:quality}.
We operationalized the seven heuristics (i.e., the 7 Cs) as described in the proposing guidelines with all its sub-heuristics wherever possible.
Some heuristics, however, were not associated with any measurement in the original publication, particularly the attribute \textit{consistent abstraction}.
We measured this by counting \textit{white-box steps}, i.e., number of steps in the scenarios of a UC that go beyond the black-box view (i.e., user-system-interaction) and describe white-box processes (i.e., system-system-interactions).
The example UC in \Cref{fig:uc} contains a highlighted white-box step.
Additionally, we relaxed the conditions for \textit{consistent grammar} to ``using active voice and simple present tense.''
Phalp et al. additionally recommend that ``adverbs, adjectives, pronouns, synonyms and negatives should be avoided''~\cite{phalp2007assessing}, which we disregarded given the already extensive variation of sentence structures appearing in the data.
An example for a grammatically inconsistent step is highlighted in \Cref{fig:uc}.

\begin{table}[hbt]
    \centering
    \caption{Quality attributes (and which heuristic (C) of the guidelines~\cite{phalp2007assessing} they belong to) per template-based use case}
    \label{tab:attributes:uc:quality}
    \begin{tabularx}{\linewidth}{p{1.5cm}lXp{0.7cm}}
    \toprule
        \textbf{Attribute} & \textbf{C} & \textbf{Description} & \textbf{Type} \\
        \midrule
        Cogent text order & 2 & Whether the sequences in the UC follow a logical order & bool \\
        End-to-end dependencies & 2 & Whether the UC describes an end-to-end interaction & bool \\
        Coherence & 3 & Percentage of sentences in a sequence that repeat a noun from the previous sentence & $[0, 1]$ \\
        White-box steps & 4 & Number of steps in a sequence that describe system-system interactions & $\mathbb{N}$ \\
        Misplaced variations & 5 & Number of steps in the main scenario that describe alternatives & $\mathbb{N}$ \\
        Misplaced main steps & 5 & Number of steps in the alternative scenario that describe main steps & $\mathbb{N}$ \\
        Sequence (main) & 5 & Whether the main scenario is structured as a numbered sequence & bool \\
        Consistent grammar & 6 & Number of steps written in simple present tense and active voice & $\mathbb{N}$ \\
        Has alternatives & 7 & Whether a description contains a separate section for extensions & bool \\
        Numbering & 7 & Whether the numbering system of alternatives is consistent with the main scenario & bool \\
        \bottomrule
    \end{tabularx}
\end{table}

We were unable to measure the following heuristics and sub-heuristics proposed by Phalp et al.
The heuristic \textit{coverage}, i.e., that a UC contains neither too little nor too much information, requires extensive domain knowledge inaccessible to the authors in a systematic and comprehensive manner for all the sampled UCs.
The sub-heuristic \textit{rational answer} of the heuristic \textit{cogent} assesses if a UC description provides ``a plausible answer to the problem''~\cite{phalp2007assessing}. 
This, again, requires domain knowledge we did not possess, similar to the sub-heuristic \textit{viable} of the heuristic \textit{consideration of alternatives} which implies that ``alternatives should be viable and make sense''~\cite{phalp2007assessing}.


\subsubsection{Automatic Extraction}

During our trial labeling, we noticed that several of the attributes derived from the guidelines~\cite{phalp2007assessing} do not sufficiently capture some properties of the encountered UC descriptions. 
For example, the number of interactions (i.e., communications between two actors) within each step (i.e., item in a sequence) varied.
Additionally, we noticed that several UC descriptions did not explicitly mention the involved actors due to the excessive use of passive voice in UC steps.
Instead, actors were often only implicit.
To improve the granularity of our measurement, we also recorded the additional quality attributes listed in \Cref{tab:attributes:uc:additional}.

\begin{table}[hbt]
    \centering
    \caption{Additional quality attributes per use case description}
    \label{tab:attributes:uc:additional}
    \begin{tabularx}{\linewidth}{p{1.5cm}Xp{0.7cm}}
    \toprule
        \textbf{Attribute} & \textbf{Description} & \textbf{Type} \\
        \midrule
        Entities & Number of entities & $\mathbb{N}$ \\
        Explicit Actors & Number of explicitly stated actors & $\mathbb{N}$ \\
        Interactions & Number of interactions (both between two entities or of an entity with itself) & $\mathbb{N}$ \\
        Consecutive interactions & Number of interactions where the actor initiating the interaction was the actor receiving the interaction in the previous step & $\mathbb{N}$ \\
        \bottomrule
    \end{tabularx}
\end{table}

To extract these attributes, we manually translated every UC into an semantically equivalent sequence diagram.
Afterwards, we parsed all of these sequence diagrams and automatically extracted the attributes listed in \Cref{tab:attributes:uc:additional} per UC.

\subsubsection{Response Variables}
\label{sec:method:data:response}

To answer RQ2, we required data that reflects the performance of the immediately subsequent SE activity of deriving a solution from the requirements.
We operationalized this construct by measuring the time that a requirement spent in the design stage (S2) in number of days.
With this measure, we approximated the direct effect that the quality of UC descriptions has on the subsequent activities.
Our assumption was that requirements of higher quality will, on average, reduce the design time, while requirements of lower quality will increase it due to clarifications and revisions.
The duration of subsequent activities is a prominent attribute to approximate the performance of an activity~\cite{frattini2024measuring} and was reliably tracked in the data available to us.


\subsection{Data Analysis}
\label{sec:method:analysis}


\subsubsection{Descriptive Statistics}
\label{sec:method:analysis:descriptive}

To answer RQ1, we generated visualizations for each non-nominal attribute listed in \Cref{tab:attributes:requirement,tab:attributes:uc:basic,tab:attributes:uc:quality,tab:attributes:uc:additional} (i.e., excluding acronym-type attributes).
We prioritized figures that show the evolution of an attribute over time aggregated per month.
For attributes where we did not observe any meaningful development over time we instead report the overall statistic summarized over the total time frame.

Important to note is that the attributes listed in \Cref{tab:attributes:requirement} are per business requirement and the attributes listed in \Cref{tab:attributes:uc:basic,tab:attributes:uc:quality,tab:attributes:uc:additional} are per UC, where one business requirement contains none, one, or multiple UCs.
To accommodate per business requirement visualization of UC-level attributes, we aggregated the values per business requirement via the mean of continuous and mode of categorical variables.

\subsubsection{Inferential Statistics}
\label{sec:method:analysis:inferential}

To answer RQ2 and RQ3, we employ a Bayesian data analysis (BDA) approach for statistical causal inference (SCI)~\cite{mcelreath2018statistical,pearl2009causality}.
While BDA is not yet commonplace in SE research~\cite{furia2019bayesian}, its principled approach within a causal framework allows dealing with confounders~\cite{furia2022applying} which is particularly relevant in our observational case study setting.
Due to limited space, this manuscript cannot serve as a proper introduction to BDA or SCI, but we refer the interested reader to appropriate textbooks~\cite{mcelreath2018statistical,pearl2009causality}, guidelines~\cite{furia2019bayesian,furia2022applying}, examples~\cite{torkar2020bayesian,frattini2025applying}, and our replication package for more detailed documentation.
Additionally, we will antedate one of the results in this method section as a running example where we investigate whether the adoption of the UC format shows an impact on the design time.
We follow the model proposed by Siebert~\cite{siebert2023applications} for SCI which builds on the Pearlean framework of causality~\cite{pearl2009causality,pearl2010causal} in three steps: modeling, identification, and estimation.

\paragraph{Modeling}

In the first step, \textit{modeling}, we make our causal assumptions explicit.
To this end, we create a directed acyclic graph (DAG) in which nodes represent variables, and edges represent assumed causal relationships between them~\cite{elwert2013graphical}.
Our causal DAG consists of three groups of variables corresponding to the three segments of our context as shown in \Cref{fig:rqs}:

\begin{enumerate}
    \item \textbf{Prior variables} (author, owner, and complexity): representing properties influencing the creation of UCs
    \item \textbf{UC variables}: properties of the UCs (as specified in \Cref{tab:attributes:requirement,tab:attributes:uc:basic,tab:attributes:uc:quality,tab:attributes:uc:additional})
    \item \textbf{Design time} (\Cref{sec:method:data:response}): representing a subsequent SE activity in which the UC is used
\end{enumerate}

\begin{figure}
    \centering
    \includegraphics[width=\linewidth]{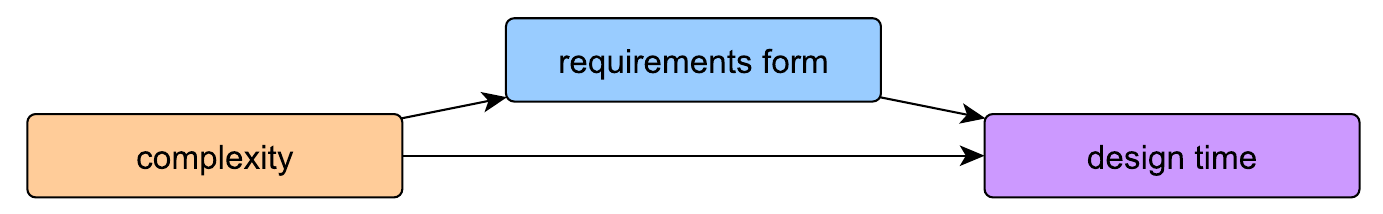}
    \caption{Causal DAG of the \textit{adoption} effect}
    \label{fig:rq2:adoption:dag}
\end{figure}

Determining the impact of UC variables on design time is in scope of RQ2, the impact of prior on UC variables is scope of RQ3.
To make the causal DAGs more manageable, we break them up into clusters of closely related sub-DAGs.
For example, \Cref{fig:rq2:adoption:dag} shows the causal DAG for the cluster representing the effect of requirements form on the design time.
Every node in the causal DAG represents a variable, where the color corresponds to the construct of \Cref{fig:rqs} that the variable operationalizes.
The two incoming edges to the purple-colored response variable \textit{design time} encode our assumption that the form of a business requirement (e.g., template-style, informal, title, user story, or no UCs at all) as well as the general \textit{complexity} of a requirement impact the design time.
Additionally, we assume that the complexity of a requirement may influence the choice of its form, e.g., since complex requirements may be preferably specified using template-style UCs.

The remaining four clusters of sub-DAGs cover the phenomena \textit{size}, \textit{solution-orientation}, \textit{coherence}, and \textit{structure} and will be reported in separate subsections in \Cref{sec:results:rq2}.
The four sub-DAGs differ in the UC variables that they connect, which are the following.

\begin{itemize}
    \item \textbf{Size}: number of use cases, total number of interactions
    \item \textbf{Solution-orientation}: end-to-end dependencies, white-box steps, location, number of explicit actors
    \item \textbf{Coherence}: coherence, cogent text order, numbers of consecutive interactions
    \item \textbf{Structure}: main scenario numbered, misplaced variations, misplaced main steps, has alternatives, numbering
\end{itemize}

\paragraph{Identification}

In the identification step, we select variables with an assumed causal impact between each other (i.e., an \textit{exposure} and an \textit{outcome}).
For RQ2, we investigate the effect of all UC variables on the SE variable \textit{design time}.
For RQ3, we investigate the effect of all prior variables on those UC variables where the analysis in scope of RQ2 confirmed a causal effect, as only these are actually relevant.

For every set of exposure and outcome variables, we identify the \textit{adjustment set}~\cite{mcelreath2018statistical}, i.e., the set of variables that need to be controlled to deconfound the effect of the exposure on the outcome~\cite{pearl2009causality}.
Applying the so-called \textit{backdoor criterion}~\cite{pearl2010causal}, we derive a statistical model from our causal model that reliably blocks confounding effects.
For example, if we aim to determine the effect of \textit{requirement form} on the \textit{design time} as shown in \Cref{fig:rq2:adoption:dag}, we need to control for \textit{complexity} as that variable is an assumed common cause of both the exposure and the outcome and, therefore, confounds the effect~\cite{cinelli2024crash}.


\paragraph{Estimation}

With the statistical model derived from our causal model via the adjustment set, we can estimate the average causal effect (ACE) via regression models.
For this, we first specify our regression formula that maps the outcome variable to the set of predictors included in the adjustment set.
At this step, we also consider more complex relationships like interactions between variables.
In our example, we assume an interaction effect between the form ($form$) and the complexity ($cplx$), i.e., we assume that the form of a requirement has different effects on the design time depending on the complexity level of that requirement.
This yields the regression formula $designtime \sim form \times cplx$.
By design, the interaction term $form \times cplx$ also implies the effects of the individual terms $form$ and $cplx$ without explicitly specifying them.

Next, we select an appropriate distribution type to represent the outcome variable based on ontological assumptions and the maximum entropy criterion~\cite{jaynes2003probability}.
All statistical models in RQ2 model the \textit{design time} variable as a zero-inflated negative binomial distribution, as the variable measured in number of days is either zero or a positive, real number with no upper bound, but an index of dispersion greater than 1~\cite{forthmann2021reliability}.
The statistical models in RQ3 regress on different outcome variables.
For example, the model investigating the effect of RE-variables on the number of \textit{white-box steps} uses a Binomial distribution, as the variable is 0 or positive and bounded by the total number of steps.
All other choices of distribution types are justified in our replication package.

With the regression formula and distribution type specified, we select prior distributions for the coefficient of each predictor~\cite{wesner2021choosing}.
We select uninformative priors (e.g., normal distributions centered around 0 with a wide standard deviation) to encode our uncertainty about the existence and strength of an effect.
Once we confirmed the eligibility of our selected priors via prior predictive checks, we train our regression models with the collected data~\cite{wesner2021choosing}.
In this process, Hamiltonian Monte Carlo Markov Chains updates the predictor coefficients~\cite{brooks2011handbook}.

\begin{figure}
    \centering
    \includegraphics[width=\linewidth]{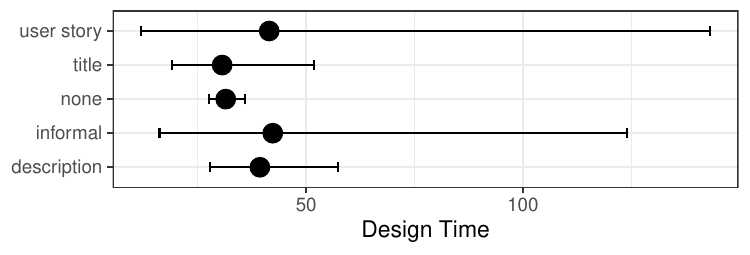}
    \caption{Marginal effect of the requirement form}
    \label{fig:rq2:adoption:form}
\end{figure}
\begin{figure}
    \centering
    \includegraphics[width=\linewidth]{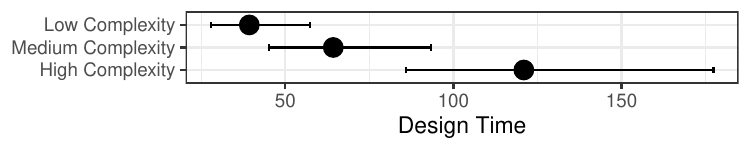}
    \caption{Marginal effect of the complexity level}
    \label{fig:rq2:adoption:cplx}
\end{figure}

Once we confirmed that the training succeeded via posterior predictive checks~\cite{mcelreath2018statistical}, we evaluate the models by plotting marginal effects of all predictors on the outcome variable.
The marginal effects simulate the effect of changing the value of the selected predictor while holding all other predictors at a constant, representative value.
This evaluation preserves all uncertainty that the model picked up in every coefficient and gives a more reliable insight into the isolated ACE of a predictor on the outcome variable.
\Cref{fig:rq2:adoption:form} shows the marginal effect of the requirement form on the design time.
The mean ACE of all forms (represented as the dot) is close to each other and the confidence intervals (CIs) overlap, showing that no form is superior by default.
However, the CI of the form \textit{none} (meaning: not following the UC approach) is the most narrow and, therefore, the most clear.
Requirements specified as \textit{user stories} or \textit{informally} exhibit the most variance on the design time.
In contrast, the marginal effect of the different levels of complexity shown in \Cref{fig:rq2:adoption:cplx} shows how more complex requirements require more design time.

\begin{figure}
    \centering
    \includegraphics[width=\linewidth]{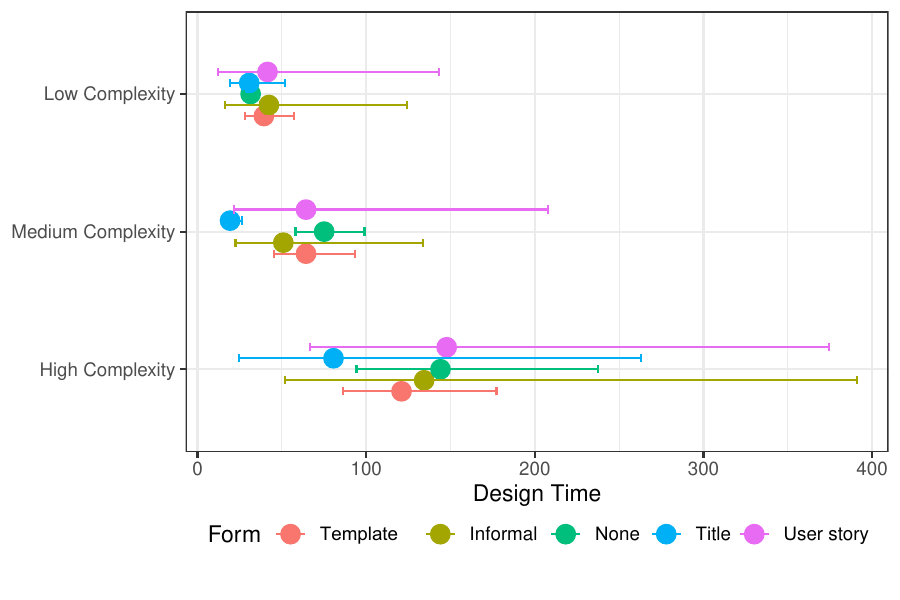}
    \caption{Conditional effect of the interaction between the complexity level and the form of a requirement}
    \label{fig:rq2:adoption:conditional}
\end{figure}

A special form of marginal effects are conditional effects, which represent the interaction between two predictors (if specified in the regression model).
\Cref{fig:rq2:adoption:conditional} shows the aforementioned conditional effect between the complexity level and requirement form.
The conditional effect shows that medium-complexity requirements have the shortest average design time when their form is just a \textit{title}.
On the other hand, proper template-style UC \textit{descriptions} exhibit the least variance for high-complexity requirements, suggesting them to be an eligible format to deal with complex requirements.

\subsection{Validation}
\label{sec:method:validation}

Finally, we shared our results with two senior managers at the case company to validate the drawn inferences against their perception.
Runeson and H{\"o}st recommend triangulating both the data source and research method to strengthen the validity of results~\cite{runeson2009guidelines}.
In our case, the two managers were involved in the initiative to adopt UCs at the case company.
Their feedback gathered in a focus group setting~\cite{kontio2008focus} allowed to appraise the observed phenomena and provided additional insights that could not be drawn from the quantitative document analysis alone.
Where applicable, we report these insights providing probable explanations for our observations in the respective results sections.

\section{Results}
\label{sec:results}

Due to space limitations, we focus on reporting meaningful and significant results.
All other analyses, figures, and coefficient tables can be found in our replication package.
All subsequently presented results are particular to the studied case and do not imply generalizability.

\subsection{Results for RQ1 (Use Case Adoption)}
\label{sec:results:rq1}

\subsubsection{UC-level Attributes}

We first visualize the UC-level attributes (i.e., \Cref{tab:attributes:uc:basic,tab:attributes:uc:quality,tab:attributes:uc:additional}).
Of the 1192 recorded UCs, 839 (70.4\%) consisted of a title, 273 (22.9\%) of a template-style UC, 44 (3.7\%) were informal, and 36 (3.0\%) were specified as user stories.
\Cref{fig:uc:informal,fig:uc:title,fig:uc:userstory,fig:uc:template} show one exemplary UC per form taken from the case.
The informal UC in \Cref{fig:uc:informal} illustrates a UC that fulfills all semantic conditions of a UC yet follows no structure.
During the validation of the results, the two senior manager hypothesized that many of UCs specified only as titles may be caused by their authors attempting to force non-functional requirements (NFRs) into a UC-format, like the deployment requirement in \mbox{\Cref{fig:uc:title}}.
Being unable to fit NFRs into a UC description, they might have resorted back to concise, high-level descriptions and labeled them UCs to comply with the initiative of their adoption.

\begin{figure}
    \centering
    \begin{tikzpicture}
        \node[draw, text width=0.97\columnwidth, align=left](req){
            \small
            \textbf{Use-case 3}: Anna and Bob are two retail postpaid customers. Bob has agreed to pay for Anna's all charges. Bob want to receive one invoice for all charges including his own charges and charges for Anna. \\
        };
    \end{tikzpicture}

    \caption{UC from the case in informal form}
    \label{fig:uc:informal}
\end{figure}
\begin{figure}
    \centering
    \begin{tikzpicture}
        \node[draw, text width=0.97\columnwidth, align=left](req){
            \small
            \textbf{Use case 1.1} Maiden Deploy of Single Site (OCS/BLR SOT19) \\
        };
    \end{tikzpicture}

    \caption{UC from the case in title form}
    \label{fig:uc:title}
\end{figure}
\begin{figure}
    \centering
    \begin{tikzpicture}
        \node[draw, text width=0.97\columnwidth, align=left](req){
            \small
            \textbf{Use Case 2}: \textit{As a} Business Config Engineer, I expect the system to automatically, create a checkpoint/tag with a default label and timestamp as a pre-step when performing a business config import \textit{so that} checkpoint can be used as a reference when triggering a business config rollback. \\
        };
    \end{tikzpicture}

    \caption{UC from the case in user story form (\textit{italics} from the original)}
    \label{fig:uc:userstory}
\end{figure}
\begin{figure}
    \centering
    \begin{tikzpicture}
        \node[draw, text width=0.97\columnwidth, align=left](req){
            \small
            \textbf{Title}: New Micro-service \\
            \vspace{0.1cm}
            \textbf{Level}: System \\
            \vspace{0.1cm}
            \textbf{Main scenario}:
            \begin{enumerate}[label=\arabic*.]
                \item A new Micro-service for Generic Batch utility is created including the creation of a new docker image, helmfile integration, pipeline establishment.
                \item Devops/SI are able to deploy new Micro-service as a separate pod/service
            \end{enumerate}
        };
    \end{tikzpicture}

    \caption{UC from the case in template form}
    \label{fig:uc:template}
\end{figure}

The remaining results refer to the 273 template-style UCs.
Almost all of these contained a title ($272/273=99.6\%$) and main scenario ($270/273=98.9\%$) like the example in \Cref{fig:uc:template}, and many contained explicit fields for actors ($128/273=46.8\%$), alternative scenarios ($156/273=57.1\%$), and additional information ($176/273=64.5\%$). 
Several UC descriptions also contained a field defining their \textit{level}, though this was mostly set to ``System'' (in $125/144=86.8\%$ of all cases) and only rarely to ``User'' ($13/144=9.0\%$), ``Business'' ($4/144=2.8\%$) or was left empty ($2/144=1.4\%$).
Pre- and postconditions were not specified often ($58/273=21.2\%$ and $31/273=11.7\%$, respectively).
The remaining fields (assumptions, solution hints, dependencies, limitations, and revision history) occurred only rarely (less than $31/273=11.3\%$).


When a business requirement followed the common structure and contained separate subsections for the business background, business requirements, and solution proposals (263 out of 273 cases), the \textit{location} of UC descriptions was either in the subsections ``Business Requirements'' ($127/263=48.2\%$) or ``Solution Proposal'' ($132/263=50.2\%$), only rarely in ``Business Background'' ($4/263=1.5\%$).

\begin{figure}
    \centering
    \includegraphics[width=\linewidth]{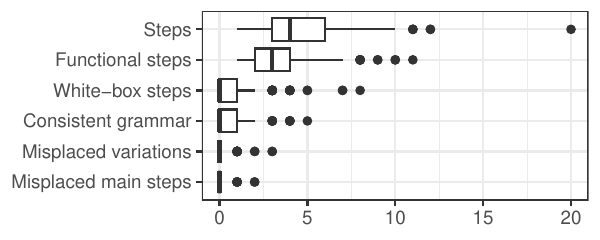}
    \caption{Distribution of steps}
    \label{fig:rq1:steps}
\end{figure}

\Cref{fig:rq1:steps} shows the distribution of steps per UC description.
While the 273 UC descriptions contained, on average, 4 steps, only 3 on average were actually functional. 
The low degree of white-box steps, misplaced variations and misplaced main steps is commendable, but the equally low number of steps with consistent grammar indicates the excessive use of passive voice to specify UC steps.

\begin{figure}
    \centering
    \includegraphics[width=\linewidth]{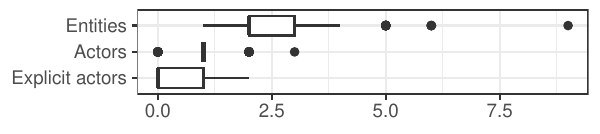}
    \caption{Distribution of entities}
    \label{fig:rq1:entities}
\end{figure}

\Cref{fig:rq1:entities} shows the distribution of entities per UC description.
Most UCs describe the interaction between two entities with an average of one being a non-system actor.
However, external actors are often just implied within the scenarios, not mentioned explicitly.

\begin{figure}
    \centering
    \includegraphics[width=\linewidth]{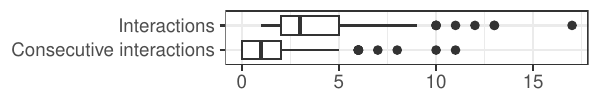}
    \caption{Distribution of interactions}
    \label{fig:rq1:interactions}
\end{figure}

\Cref{fig:rq1:interactions} shows the distribution of interactions per UC description.
The average number of interactions (3) is equal to the average number of functional steps seen in \Cref{fig:rq1:steps}, which indicates that most steps in UC descriptions cover only one interaction (i.e., either a user request or a system response), not both.
Also, the number of consecutive interactions is considerably lower than the number of interactions on average.

\subsubsection{Business Requirement-level Attributes}

\begin{figure}
    \centering
    \includegraphics[width=\linewidth]{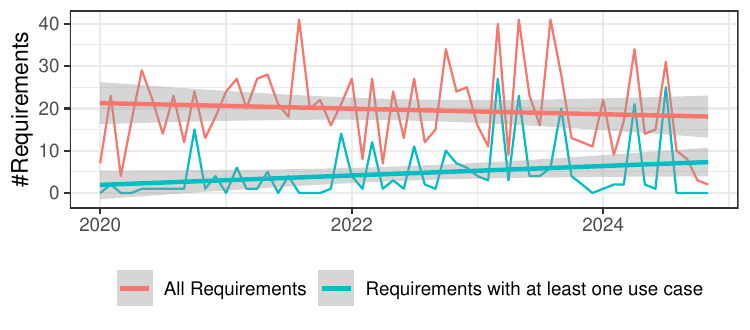}
    \caption{Number of requirements per month}
    \label{fig:rq1:uca:absolute}
\end{figure}

In this section, we report the results on business requirements level.
\Cref{fig:rq1:uca:absolute} visualizes the absolute number of business requirements (in red) as well as the absolute number of requirements containing at least one UC (in cyan) over the surveyed time frame.
The trend lines show that while the absolute number of requirements produced each month slightly declined, the absolute number of requirements containing at least on UC rose.
The relative level of business requirements adopting the UC approach increased from roughly 10\% in the start of 2020 to roughly 30\% at the end of 2024.
On the other hand, the average number of UCs per business requirement decreased from 5 to 3 over this time span.

\begin{figure}
    \centering
    \includegraphics[width=\linewidth]{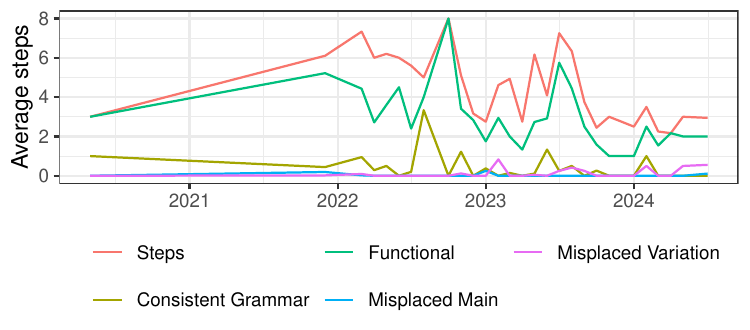}
    \caption{Evolution of number of steps}
    \label{fig:rq1:steps:evolution}
\end{figure}

\Cref{fig:rq1:steps:evolution} shows the evolution of the average number of steps per UC over time.
The average number of steps increased up to 2022 and stabilized again afterwards.
The evolution of the number of interactions and entities shows a similar trend.
The two managers involved in the validation explained that the phase between early 2022 and mid 2023 coincided both with an active promotion of the UC-driven approach at the company and a major transformation of the architecture of the developed product.
These two factors likely explain the spike in UC size.

\begin{figure}
    \centering
    \includegraphics[width=\linewidth]{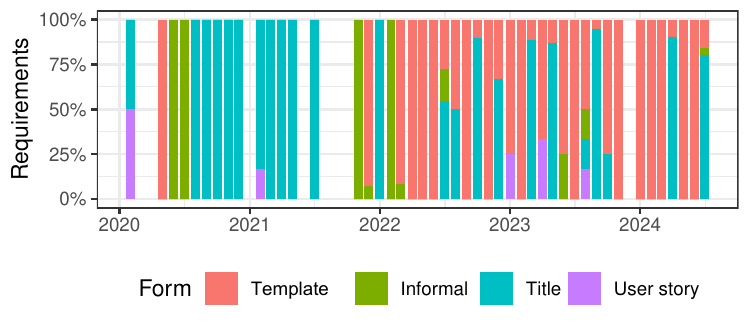}
    \caption{Evolution of UC forms}
    \label{fig:rq1:form:evolution}
\end{figure}

Finally, \Cref{fig:rq1:form:evolution} shows the evolution of the distribution of UC forms over time.
While up to 2022 most UCs were specified using only titles, with the beginning of 2022 proper UC descriptions became the dominant form.
This, again, coincides with the active promotion of the UC-driven approach starting 2022.

In summary, we observed that the case company adopted a UC-based approach to requirements specification, though the usage deviates from textbook recommendations of UC specification.
Active, top-down promotion of the approach stabilized the adoption of proper template-style UCs for about 30\% of all new business requirements.

\subsection{Results for RQ2 (Use Case Quality Impact)}
\label{sec:results:rq2}

The following subsections report the results for RQ2 per cluster as mentioned in \Cref{sec:method:analysis:inferential}.
Note that while the running example in \Cref{sec:method:analysis:inferential} inferred results by considering all business requirements, the following analyses are limited to requirements containing template-style UCs since we want to determine the impact of their quality.

\subsubsection{Size}

Investigating the hypothesis that the size of requirements (in terms of \textit{number of UCs} and \textit{total number of interactions}) has an effect yielded no meaningful results.
The design time was not influenced significantly by either of the two variables, though high-complexity requirements showed a greater variance in design time for greater number of UCs.

\subsubsection{Solution-orientation}

\begin{figure}
    \centering
    \includegraphics[width=\linewidth]{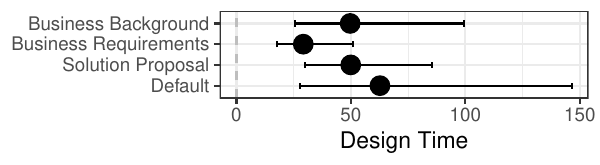}
    \caption{Marginal effect of UC location on design time}
    \label{fig:rq2:solution:location}
\end{figure}
\begin{figure}
    \centering
    \includegraphics[width=\linewidth]{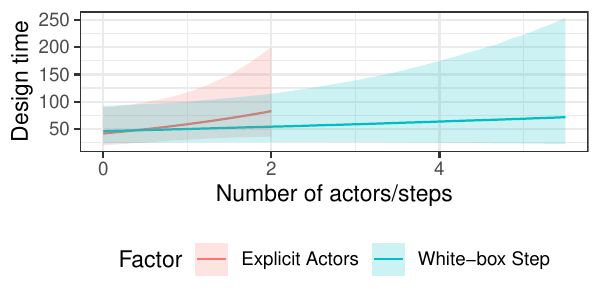}
    \caption{Marginal effect of white-box steps and explicit actors}
    \label{fig:rq2:solution:whitebox:actors}
\end{figure}

\Cref{fig:rq2:solution:location} shows the marginal effect of the location of UCs on the design time.
The effects are very uncertain (as all CIs overlap), but the mean values (represented by the dots) show that UCs specified in the ``Business Requirements'' section of a requirement require less design time than UCs specified in the ``Solution Proposal.''
This is additionally supported by the marginal effect of the number of white-box steps shown in \Cref{fig:rq2:solution:whitebox:actors} (in cyan).
The more white-box steps a UC contains, the more days than average a requirement spent in the design phase.
This hints at the negative effect of solution-orientation which we will discuss in \Cref{sec:implications}.
On the other hand, the increasing number of explicit actors is also associated with a longer design time as shown in \Cref{fig:rq2:solution:whitebox:actors} (in red).

\subsubsection{Coherence}

\begin{figure}
    \centering
    \includegraphics[width=\linewidth]{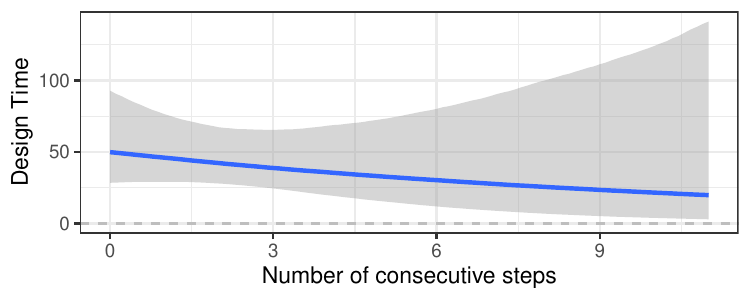}
    \caption{Marginal effect of consecutive steps}
    \label{fig:rq2:coherence:consecutive}
\end{figure}

When investigating the variables related to the coherence of a UC description, we observed a notable effect of the number of \textit{consecutive steps} on the design time seen in \Cref{fig:rq2:coherence:consecutive}.
UCs with more consecutive steps, i.e., scenarios where interactions pick up where the previous left of, decreased the design time.
The effect is, however, non-significant due to its wide CI (grey ribbon in \Cref{fig:rq2:coherence:consecutive}).

\subsubsection{Structure}

\begin{figure}
    \centering
    \includegraphics[width=\linewidth]{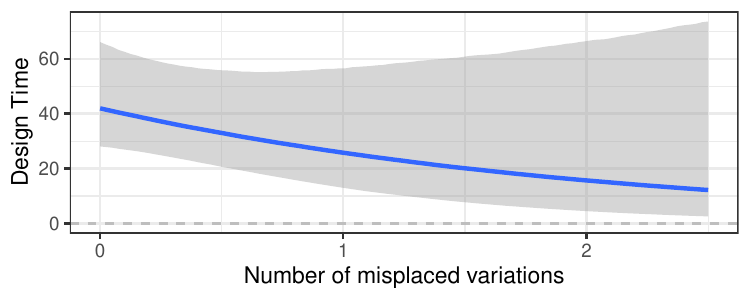}
    \caption{Marginal effect of misplaced variations}
    \label{fig:rq2:structure:variations}
\end{figure}

We observed only one notable effect by the structure variables of the UC descriptions.
The number of \textit{misplaced variations} has a considerable reducing effect on the design time as seen in \Cref{fig:rq2:structure:variations}.
This means, the more variations are present in the main scenario, the lower the design time on average.

\subsection{Results for RQ3 (Factors influencing Use Case Quality)}
\label{sec:results:rq3}

We investigate the effects of RE-variables on the four most significant UC quality factors: number of use cases, number of white-box steps, number of misplaced variations, and number of explicit actors.
Given their impact on the design time, understanding which RE-factors influence them yields the greatest control to the case company.

\subsubsection{Number of use cases}

None of the available predictors of this outcome variable showed a significant effect.
The analysis merely yielded that medium- and high-complexity requirements, on average, contained 3 UCs while low-complexity requirements contained 2.

\subsubsection{White-box steps}

\begin{figure}
    \centering
    \includegraphics[width=\linewidth]{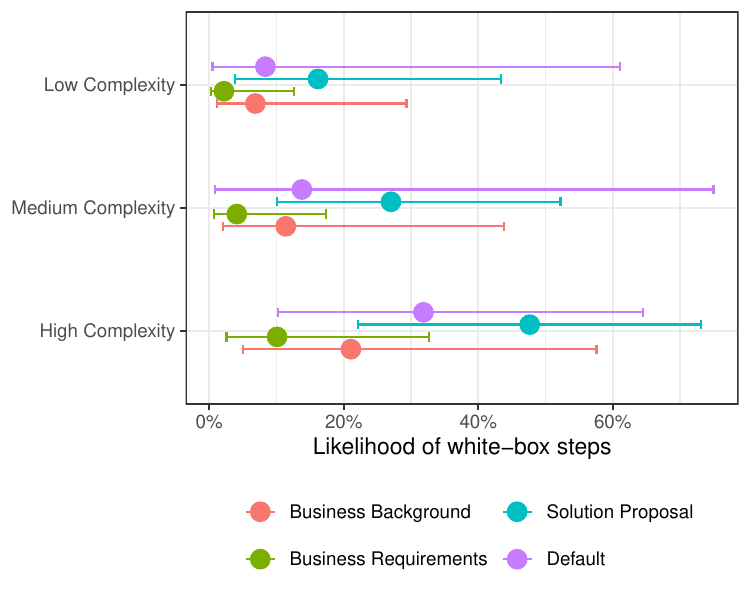}
    \caption{Conditional effect of location and complexity}
    \label{fig:rq3:stepsbeyond:location:cplx}
\end{figure}

Both the location of a UC within a structured requirement as well as the complexity have an influence on the number of white-box steps in its description.
\Cref{fig:rq3:stepsbeyond:location:cplx} shows how UCs located in the ``Business Requirements'' section are much less likely to contain white-box steps on all levels of complexity.
High-complexity requirements are, in general, more prone to contain white-box steps.

\subsubsection{Misplaced variations and explicit actors}

We did not detect any RE-variable with a significant effect on these two UC quality variables.
Worth noting is also that in none of the four models we detected a significant impact of the UC owner or author on the respective UC quality variable.

\section{Discussion}
\label{sec:discussion}

In the following, we summarize insights drawn from the data (\Cref{sec:discussion:general}) and present implications derived from them (\Cref{sec:discussion:specific}).
While the insights are supported by the empirical data, the implications were devised together with the two senior managers based on this data.
These conclusions may not be generalizable, yet we present them as exemplary actionable insights potentially relevant for comparable cases.
Finally, we discuss the threats to validity to the conclusions drawn from our study following the categorization by Wohlin et al.~\cite{wohlin2012experimentation}.

\subsection{Insights into Use Case Adoption and Quality}
\label{sec:discussion:general}

While related work presents UC descriptions as a prominent format to specify functional requirements~\cite{anda2003empirical,anda2009investigation,tiwari2015systematic}, the evolution of its adoption at our observed case depended on the active promotion of the approach by senior management.
Beyond that, the natural adoption of UC-driven requirements deviates significantly from proposed guidelines~\cite{phalp2007assessing}.
This entails a wide variety of used fields, the misuse of UCs to specify system-internal rather than user-system interactions, the lack of explicitly defined actors, and inconsistent grammar through the use of passive voice.

The adoption of UC descriptions for requirements specification does not automatically lower the design time in our case.
However, \Cref{fig:rq2:adoption:conditional} shows that UC descriptions may be viable to capture highly complex requirements.
This underlines that merely introducing UCs may not improve the SE process in itself, but has potential to deal with complex requirements.

When investigating how the quality of template-style UC descriptions affects subsequent SE activities in the scope of RQ2, the phenomenon of solution-orientation stood out as a significant factor.
Solution-oriented UC descriptions that describe system-internal interactions from a white-box perspective tend to delay the subsequent design phase, presumably since premature design decisions need to be revisited, explained, and revised.
The results, again, emphasize that the adoption of a UC-driven approach to requirements specification does not preempt solution-oriented requirements by design.
While the RE literature quite explicitly declares the primary purpose of UCs to describe user-system interactions~\cite{kulak2001use}, the approach was naturally misused for system-system interactions.

On the other hand, some rules of UC quality guidelines like the recommendation to keep all variations out of the main scenario of a UC description~\cite{phalp2007assessing} were met with contrary results in our studied case.
As seen in \Cref{fig:rq2:structure:variations}, the increased conflation of the main scenario with alternative steps actually benefited the duration of the subsequent design time.
The complexity of connecting steps of the main scenario to alternatives in the extensions might be an explanation for this.

The investigation in scope of RQ3 revealed little insights into factors that affect UC quality. 
Only the location of a UC within a requirement, i.e., declaring its application to specify either a requirement or solution proposal, is a valuable indicator of whether the UC is solution-oriented or not.

\subsection{Implications for the Case Company}
\label{sec:discussion:specific}

Firstly, the managers confirmed the company's commitment to the UC-driven requirements specification approach.
While the adoption of UC descriptions did not lower the design time by default, it enabled a transparent and fine-grained discussion and analysis of requirements specifications.
Conducting these discussions and analyses based on empirical studies instilled confidence in an iterative and evidence-based improvement of their RE approach.

Secondly, the level of abstraction of UCs---i.e., whether they specify requirements or potential solutions---will be assessed more carefully in the future.
While it may be infeasible to inhibit the natural tendency of requirements authors to misuse UCs to specify potential solutions rather than actual requirements, the impact of white-box steps and UC location on design time shown in \Cref{fig:rq2:solution:location,fig:rq2:solution:whitebox:actors} was perceived as significant.
Identifying solution-oriented UCs and anticipating that they change is expected to be a valuable strategy to minimize design time overrun in the future.
To identify these solution-oriented UCs, the company will use both the location of the UC (i.e., whether it is located in the ``Solution Proposal''-section of the business requirements) as well as the depth of interactions in sequence diagrams generated from UCs.

Thirdly, the managers encourage the use of coherent UC steps given their positive effect on design time shown in \Cref{fig:rq2:coherence:consecutive}.
Additionally, incoherent steps were understood as a heuristic to detect missing steps, as the lack of a repeated noun between two steps may stem from an omitted intermediate step.

\subsection{Threats to Validity}
\label{sec:discussion:threats}

\subsubsection{Threats to internal Validity}

Most importantly, the validity of conclusions drawn in the scope of RQ2 and RQ3 may be threatened by unobserved confounders.
In our study, we were limited to observable and accessible variables in scope of our collaboration.
However, there may be several additional factors that could have influenced the observed relationships.
For example, the motivation and skill of the involved stakeholders, political decision, or the pressure under which certain requirements of higher importance were addressed could all influence the observed effects.
The immeasurability of these mostly latent variables additionally complicates their inclusion in a quantitative, statistical model.
While we cannot mitigate this threat to internal validity, we address it via our employed data analysis approach that makes all of our causal assumptions explicit.
Disclosing all of our causal assumptions in the form of DAGs allows proposing competing assumptions and empirically comparing them~\cite{furia2019bayesian,mcelreath2018statistical}.

Finally, the validation of our results in the focus group setting described in \Cref{sec:method:validation} may be affected by a bias of the involved participants, as they oversaw the adoption of UCs at the case company.
We mitigated this threat to the degree possible by steering the conversation toward the explanation of unclear phenomena without enticing judgment.

\subsubsection{Threats to construct Validity}

Additionally, several operationalizations of concepts of interest are of unclear construct validity.
The degree to which the quality guidelines by Phalp et al.~\cite{phalp2007assessing} actually represent UC quality is questionable given the lack of empirical evidence, yet we hope to contribute to this with our study.
On the other hand, our response variable \textit{design time} is an admittedly simplistic operationalization representing the performance of the subsequent solution design phase.
Within the scope of our study, it was the variable that most reliably reflected the response concept of interest.
Still, triangulation by additionally considering the impact of other attributes like solution specification correctness are necessary to comprehensively study the effect of UC quality in the case study more holistically~\cite{femmer2015s,frattini2024measuring}, but this was beyond our possible data collection.
Finally, attempting to quantify phenomena about quality may generally be disputable.
We consider this threat partially mitigated as the complementary qualitative feedback during the focus group validation deemed all observed quantitative effects feasible.
While this cannot be construed as systematic evidence, it at least strengthened the confidence in the validity of the observed effects.

\subsubsection{Threats to conclusion Validity}

The complex approach of BDA for data analysis offers more in-depth inferences~\cite{mcelreath2018statistical} at the expense of potential threats to conclusion validity.
In this study, we did not consider more complex hierarchical models or distribution types which might fit the data better.
However, by design, BDA aims to make all subjective decisions like the choice of a distribution type or prior distributions explicit.
We documented all these decisions pertaining to our data analysis in our replication package for review, reuse, and improvement.

\subsubsection{Threats to external Validity}

This study describes a case study, which makes no claim at generalizability by design~\cite{wohlin2021case}.
The results are particular to the described case and cannot be transferred into other contexts without careful consideration.
Rather, we view our results as a starting point for more general recommendations about whether and how to adopt a UC-driven approach.
Achieving these general recommendations requires replication of our study in different contexts.


\subsection{Future Work}

This descriptive and explanatory case study lays the foundation for effective follow-up research.
From a researcher-perspective, we plan to iteratively improve our causal models and capture the latent variables that potentially confound the observed causal effects.
Operationalizing and approximately measuring the skill of the involved stakeholders or the pressure applied based on requirements' importance may improve the insights into the causal effects.
Furthermore, extending the considered response variables to other attributes (e.g., the correctness of the design phase measured via number of defects) as well as other activities (e.g., how UC descriptions affect implementation and testing) would provide a more comprehensive insight into the effects of UC quality.

From a practical perspective, we plan to utilize the obtained knowledge and---jointly with the case company---act on the factors that influence the performance of subsequent SE-activities.
In a study considered an \textit{improving case study} or \textit{action research} by Robson~\cite{robson2002real}, we would support the case company in addressing solution-oriented UCs already during the RE phase to observe whether its negative effect on the design time can be mitigated.

\section{Conclusion}
\label{sec:conclusion}

UC descriptions are a potent format for specifying requirements.
Despite a plethora of quality guidelines, the adoption of UCs in practice is understudied.
With this case study, we provide insights into the adoption of a UC-driven requirements specification approach over five years at a globally distributed case company.
The results show that the practical adoption of UCs deviates from textbook recommendations and that issues like solution-orientation affected the case throughout.
We hope to inspire empirical research on the adoption and application of RE specification formats but also to steer requirements quality research into a relevant direction, focusing on factors of UC quality that show an impact in practice.

\section*{Acknowledgment}
This work was supported by the KKS foundation through the S.E.R.T. Research Profile project at Blekinge Institute of Technology.
We are deeply grateful to Parisa Yousefi and Charlotte Ljungman for enabling and supporting the collaboration.
Finally, we owe great thanks to Daniel Mendez for his invaluable feedback.

\bibliographystyle{IEEEtran}
\bibliography{material/references}

\end{document}